\documentclass[12pt]{article}
\usepackage{graphicx}
\usepackage{atlasphysics}
\usepackage{subfigure}
\usepackage[table]{xcolor}
\usepackage{hyperref}
\usepackage{array}
\newcolumntype{x}[1]{>{\centering\arraybackslash\hspace{0pt}}p{#1}}

\newcommand\pubnumber{IFIC/17-01}
\newcommand\pubdate{\today}

\def\ific{IFIC (UVEG/CSIC) Valencia  \\
C./Catedratico Jose Beltran 2, Paterna, E-46980, Spain}

\def\Title#1{\begin{center} {\Large #1 } \end{center}}
\def\Author#1{\begin{center}{ \sc #1} \end{center}}
\def\Address#1{\begin{center}{ \it #1} \end{center}}

\newcommand\pubblock{\rightline{\begin{tabular}{l} \pubnumber\\
         \pubdate  \end{tabular}}}
\newenvironment{Abstract}{\begin{quotation}  }{\end{quotation}}
\newenvironment{Presented}{\begin{quotation} \begin{center} 
             PRESENTED AT\end{center}\bigskip 
      \begin{center}\begin{large}}{\end{large}\end{center} \end{quotation}}
\def\Acknowledgements{\bigskip  \bigskip \begin{center} \begin{large}
             \bf ACKNOWLEDGEMENTS \end{large}\end{center}}

\def\beq{\begin{equation}}
\def\eeq#1{\label{#1}\end{equation}}
\def\eeqn{\end{equation}}

\def\beqa{\begin{eqnarray}}
\def\eeqa#1{\label{#1}\end{eqnarray}}
\def\eeqan{\end{eqnarray}}

\def\sst{\scriptscriptstyle}

\let\bar=\overbar

\def\Dslash{\not{\hbox{\kern-4pt $D$}}}
\def\dslash{\not{\hbox{\kern-2pt $\del$}}}

\def\msb{{\bar{\ssstyle M \kern -1pt S}}}

\begin{document}
\begin{titlepage}
\pubblock

\vfill
\Title{Top physics beyond the LHC}
\vfill
\Author{Marcel Vos}
\Address{\ific}
\vfill
\begin{Abstract}
Several proposals exist for energy-frontier facilities  
after the HL-LHC. In this contribution I review the potential of these 
facilities to perform key measurements of top quark properties and interacions.
Top quark precision physics at a lepton collider provides excellent
opportunities for the determination of the top quark mass and 
its couplings to neutral electro-weak gauge bosons.
Measurements at very high center-of-mass energy at a new hadron collider 
likely offer the ultimate precision on the QCD interaction of the top quark
and its coupling to the Higgs boson. The combination of a lepton and a
hadron collider can improve the precision of all these measurements
by two orders of magnitude with respect to the 
current state-of-the-art and by an order of magnitude compared
to the precision envisaged after 3~\iab{} at the LHC.
\end{Abstract}
\vfill
\begin{Presented}
TOP2016\\
Olomouc, Czech Republic, September 19-23, 2016
\end{Presented}
\vfill
\end{titlepage}
\def\thefootnote{\fnsymbol{footnote}}
\setcounter{footnote}{0}

\section{Introduction}

In the early 70s, when the top quark was postulated~\cite{Kobayashi:1973fv}, 
energy-frontier colliders were relatively cheap. A rich programme around the world gave
rise to a flurry of discoveries that established the Standard Model of
particle physics. In 1995 the observation of the top quark was announced by the 
experiments operating at the Tevatron~\cite{Abe:1995hr,Abachi:1995iq}. 
At the same time $e^+e^-$ colliders at SLAC and CERN could infer the top quark mass from 
precision measurements. Today, the Large Hadron Collider (LHC) operates at 
the luminosity and center-of-mass energy it was designed for. 
It is the only energy-frontier facility in the world and will 
maintain that position for a good part of its lifetime. 

With the discovery of a boson~\cite{Aad:2012tfa,Chatrchyan:2012xdj} compatible with 
the predictions for the SM Higgs boson~\cite{Khachatryan:2016vau}, the theory of 
the electro-weak and strong nuclear interactions is complete. Credible
extensions of the Standard Model are under pressure from the null result
of a large variety of searches.
In this situation there is no obvious target for the next facility of 
particle physics. We enter an era of exploration, where precise and sensitive
measurements of known processes may be the best opportunity to reveal hints 
of the high-scale physics that lies beyond the Standard Model.

An in-depth scrutiny of top quark production at colliders may uncover 
such evidence. A precise determination of the top quark mass 
(together with a precise measurement of the W-boson mass) tests the
internal inconsistency of the SM. Precise measurements of the QCD or EW 
interactions of the top quark have exquisite sensitivity to broad 
classes of extensions of the SM. A direct and
precise measurement of the Yukawa coupling of the top quark tests 
the Higgs mechanism where its coupling is strongest. Searches for 
flavour-changing-neutral-current interactions involving top quarks
target yet another sector of possible extensions of the SM. 

In this contribution I explore the progress new facilities may bring 
to top quark physics in the post-LHC era. The expectations for
the remainder of the LHC programme and its luminosity upgrade
are covered in another contribution to these proceedings~\cite{azzi}.

In Section~\ref{sec:projects} the existing projects for energy-frontier 
facilities are briefly introduced. Sections~\ref{sec:productionhad}
and~\ref{sec:productionlep} provide necessary background information 
about top quark production at hadron and lepton colliders. 
In Sections~\ref{sec:mass} through~\ref{sec:ew} the prospects for 
a number of key measurements in top physics are presented. Finally,
Section~\ref{sec:summary} presents a summary and outlook.

\newpage

\section{Future collider projects}
\label{sec:projects}

Accelerating technology has made important progress over the last decades. Where the SLAC linear collider 
was built with 17 MV/m cavities, the superconducting cavities for the ILC and XFEL reach 
accelerating gradients of 35 MV/m~\cite{Baer:2013cma}. The proof-of-principle for drive-beam acceleration 
has been demonstrated, with a gradient of up to 100 MV/m~\cite{Aicheler:2012bya}, enabling
multi-\tev{} operation in a relatively compact machine.
The last decades have also witnessed steady progress in magnet technology: LHC dipoles produce a bending field of 8 T, 
a factor two stronger than the magnets used at the Tevatron. Projects for very large hadron colliders, with an energy reach 
that exceeds that of the LHC by up to a factor 7, rely on the development of 16 T dipole magnets.

Even after succesfull conclusion of the R\&D phase, the lead time for energy frontier facilities 
(the technical design, political negotiations, construction and commissioning) is measured in decades. 
If a new facility is to be  operational by the end of the LHC programme, 
construction must start in the early 2020s.
The large cost of the next facility of this scale requires 
coordination at a global level. After the demise of the domestic collider 
programme in the US and with the long-term commitment of Europe to the LHC
programme, Asia is definitely a powerful partner, and possibly the host,
of any new large-scale facility.
In Table~\ref{tab:projects} the energy-frontier collider projects are listed.

\begin{table}[htp!]
\caption{Projects for energy-frontier facilities. Integrated luminosities correspond to different running times (typically ten years per energy point) and are subject to large uncertainties. The last column provides a reference to the most recent design report. References to detailed running scenarios are given where available. All $e^+e^-$ colliders envisage a brief period of running at energies close to the top quark pair production threshold, that is not listed here.}
\label{tab:projects}
\begin{tabular}{lccccc} \hline
\multicolumn{6}{c}{\bf energy-frontier lepton colliders} \\ 
               &      &       & $\sqrt{s} $  & $\int \cal{L} $ &   \\ 
 project       & host & type  & [TeV] & [\iab] &  status \\ \hline
        &         &        &         &            &                                     \\ 
ILC                       & Japan & linear  $e^+e^-$ & 0.25  &  0.5 (2)                  & TDR 2013~\cite{Baer:2013cma} \\ 
                          &       &        & 0.5  &  0.5 (4)                 & staging~\cite{Fujii:2015jha,Barklow:2015tja} \\
        &         &        &         &            &                                     \\ 
CLIC                      & CERN  &linear $e^+e^-$  &  0.38 & 0.5 & CDR 2012~\cite{Aicheler:2012bya} \\ 
                          &       &            & 1.5,3 & 1.5,2 & staging~\cite{CLIC:2016zwp}   \\        
       &         &        &         &            &                                     \\ 
CEPC                      & China & circular $e^+e^-$ &   0.25   &   5                  & CDR 2017~\cite{CEPC-SPPCStudyGroup:2015csa} \\
       &         &        &         &            &                                     \\ 
FCCee                     & CERN & circular $e^+e^-$  & 0.25/0.36   &  10/2.6          & CDR $<$ 2019~\cite{Gomez-Ceballos:2013zzn} \\ 
       &         &        &         &            &                                     \\ 
$\mu$ collider             & FNAL? &  racetrack $\mu^+\mu^-$  & 0.125-3 &  -         & R\&D ~\cite{Alexahin:2013ojp}       \\
        &         &        &         &            &                                     \\ \hline
\multicolumn{6}{c}{\bf very-high-energy pp colliders} \\
        &         &         &  $\sqrt{s} $  & $\int \cal{L} $  &     \\ 
project & host    &  tunnel & [\TeV]  &  [\iab]     & reference   \\ \hline 
       &         &        &         &            &                                     \\ 
VLHC     & CERN   & LEP/LHC & 25    &            &   -   \\
       &         &        &         &            &                                     \\ 
FCChh     &  CERN  &  new   &  100    & 20         & ~\cite{Arkani-Hamed:2015vfh,Golling:2016gvc,Contino:2016spe,Mangano:2016jyj}\\
       &         &        &         &            &                                     \\ 
SPPC     & China   & new    &  70-140 &  3         & ~\cite{CEPC-SPPCStudyGroup:2015csa} \\ 
         &         &        &         &            &                                     \\ \hline
\end{tabular}
\end{table}
                         
The complementarity of the lepton collider projects is clear: the circular machines can provide unrivalled luminosity at low energy, 
but have limited energy reach. For the maximum circumference envisaged (100 km) the top quark pair production threshold can be 
reached~\cite{Gomez-Ceballos:2013zzn}. For $e^+e^-$ collisions at still higher center-of-mass energy linear colliders are the only 
viable solution. The linear colliders are the most mature projects. The ILC has finalized the technical design~\cite{Baer:2013cma} 
and entered the phase of negotiations between the envisaged host (Japan) and international partners. It is followed closely by CLIC,
that has prepared an extensive conceptual design report~\cite{Aicheler:2012bya}. 

An upgrade of the LHC dipoles with 16~T bending magnets in the 
existing LHC tunnel could roughly double the energy reach. The FCChh and SPPC projects are much more ambitious: a new, circular $pp$ 
collider with a size several times that of the LHC should reach up to 100~\tev. Physics and detector studies for the latter two 
machines are ramping up rapidly. A complete CDR is expected before the update of the European Strategy in 2019.

\newpage

\section{Top quark production at future hadron colliders}
\label{sec:productionhad}

\begin{figure}[h!]
\centering
\begin{subfigure}[hadron colliders]{
\centering
\includegraphics[width=0.44\textwidth]{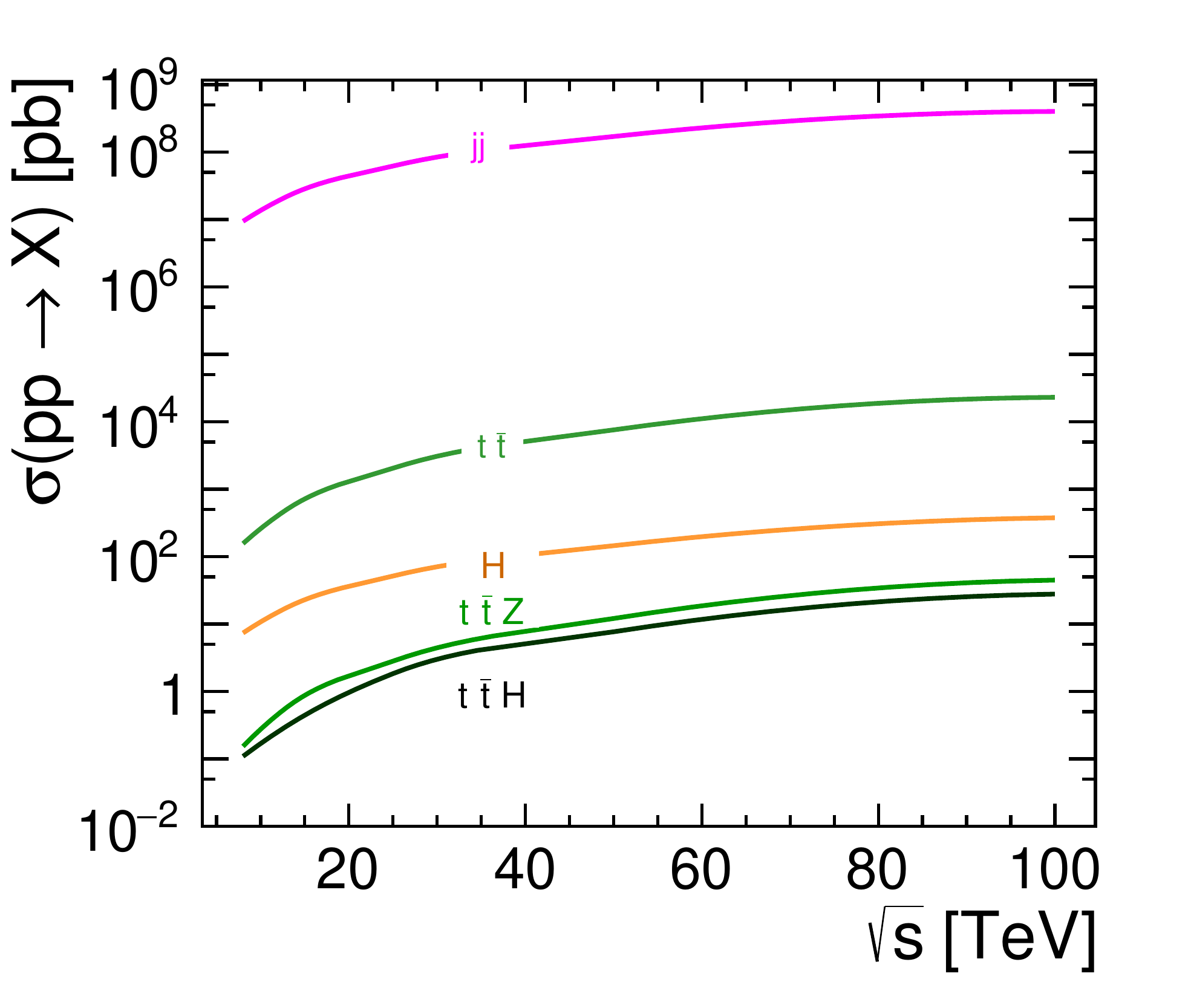}}
\end{subfigure}%
~
\begin{subfigure}[lepton colliders]{
\centering
\includegraphics[width=0.44\textwidth]{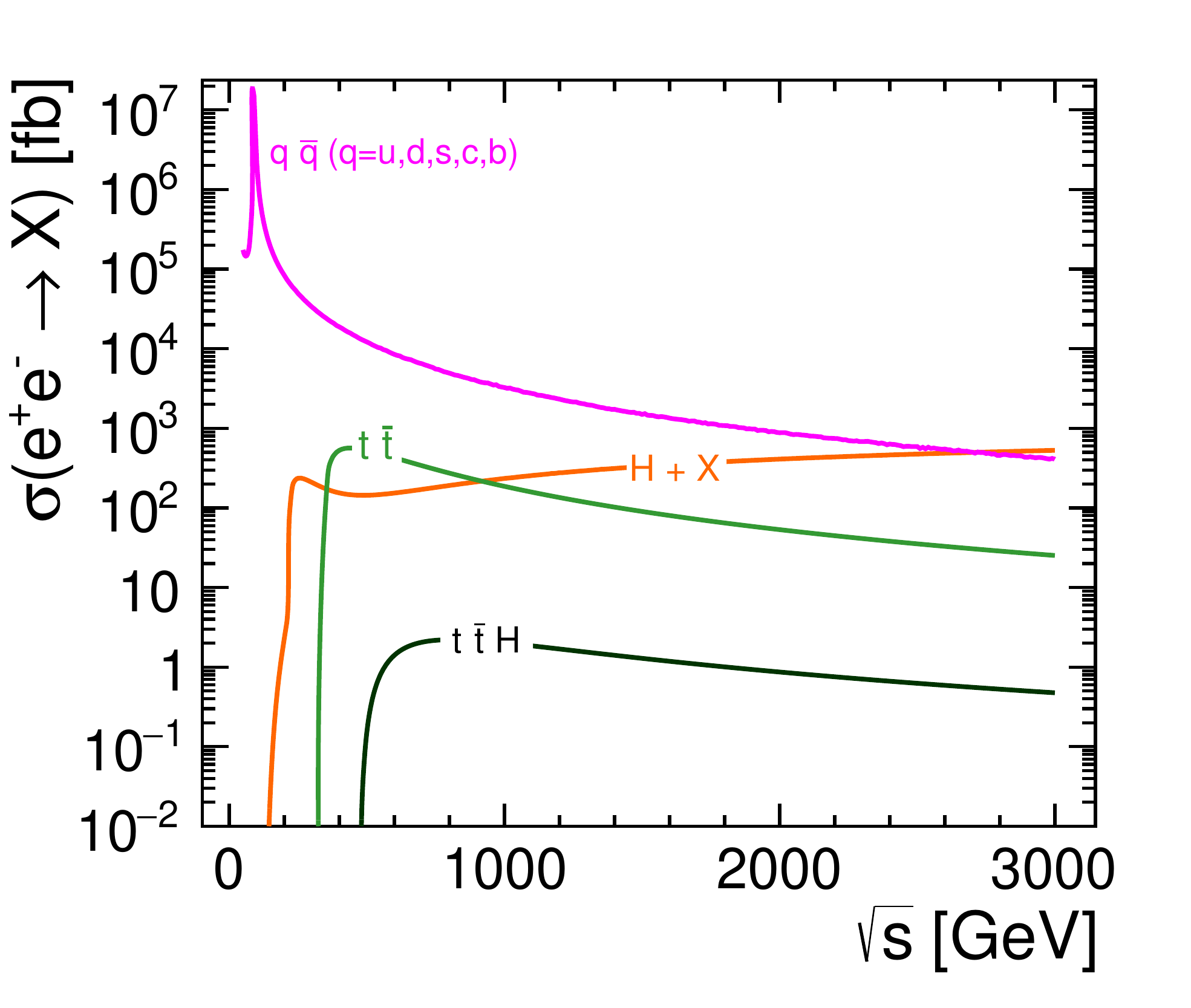}}
\end{subfigure}
\caption{The cross sections for SM processes as a function of center-of-mass energy. In panel (a) the
predictions are given for a proton-proton collider with 8~\tev{} $ < \sqrt{s} < $ 100~\tev. 
In panel (b) the unpolarized production cross section for $e^+e^-$
collisions is shown in for center-of-mass energies up to 3~\tev.}
\label{fig:xsecs}
\end{figure}

High-energy hadron colliders produce copious samples of top quarks through
QCD pair production and EW single top production. Where the Tevatron collected a 
a sample of tens of thousands of top quark pairs, the LHC has produced tens
of millions. With the HL-LHC this number will increase by two further
orders of magnitude. At a higher-energy machine several further orders
of magnitude can be added. An overview of 
cross sections for Standard Model process at a 100~\tev{} pp collider 
is given in Fig.~\ref{fig:xsecs} (a). A proton collider with a center-of-mass energy 
of 100~\tev{} and an integrated luminosity of order 10~\iab{} will
produce an astonishing sample of $10^{12}$
top quark pairs. The increase in rates is even more impressive in the 
high-mass tail and in associated production of
top quarks with gauge bosons or the Higgs boson. 

The abundant production of highly boosted top quarks 
challenges detectors and reconstruction techniques. 
The decay products from a 5~\tev{} top quark are typically collimated in 
an area with $\Delta R <$ 0.1. Traditional signatures, such as
isolated leptons and jet multiplity are expected
to be less distinctive than at the LHC. 
The detector granularity needed to resolve jet substructure is 
challenging for in particular of the (hadronic) calorimeter system. 
Selection of final states with (boosted) top quarks is therefore far 
from trivial. Several authors have explored approaches to deal with hyperboosted 
top quarks, using substructure analysis~\cite{Golling:2016gvc}, substructure 
analysis limited to charged particle tracks~\cite{Larkoski:2015yqa}, or even 
the lepton-in-jet signature~\cite{Aguilar-Saavedra:2014iga}. 

As statistical uncertainties become irrelevant even for processes
that are rare today, ``one of the key obstacles in precision
measurements at hadron colliders [...] is the intrinsic difficulty 
in performing accurate absolute rate predictions”~\cite{Mangano:2016jyj}. 
Even today, theory uncertainties exceed the experimental ones for 
the inclusive $t\bar{t}$ production cross section. The most precise 
measurements by ATLAS and CMS have a precision of approximately 4\%,
(the statistical uncertainty contributing only 0.1\%), whereas the theory 
uncertainty adds up to approximately 5-6\%~\cite{Czakon:2013goa}.
The ultimate potential of the top physics programme at hadron colliders 
therefore depends crucially on the ability to reduce systematic 
uncertainties.

\section{Top quark production at future lepton colliders}
\label{sec:productionlep}

Lepton colliders present a very different environment, where top quarks,
and especially background events, are produced at a more modest pace. 
The cross sections for important SM processes
are shown in Fig.~\ref{fig:xsecs} (b). At a center-of-mass energy of 500~\gev{}, 
close to the maximum rate, the unpolarized cross section is approximately 600~fb, 
increasing to 1~pb for polarized beams. Even the ILC luminosity upgrade 
will not produce more than a few million top quark pairs. 
At higher energy the cross section for the s-channel process drops rapidly 
(more rapidly than the luminosity increase expected at linear colliders) and 
production is even less abundant. At lepton colliders cross sections are {\em democratic}. 
Top quark pairs are the dominant source of 6-fermion final states, so that no trigger
is required and measurements can be very inclusive, with virtually no background.

The key to precision physics at lepton 
colliders is the possiblity to compare per-mil level measurements to
equally precise predictions. There is no PDF uncertainty and QCD scale 
uncertainties are of the order of several per mil. The beam energy, luminosity 
and polarization can be controlled to 0.1\%.  
From existing full-simulation studies~\cite{Baer:2013cma,Linssen:2012hp,Vos:2016tof} 
it seems that lepton collider experiments may indeed control the experimental systematics 
in top quark studies to the required level. However, a detailed study of signal modelling 
and acceptance uncertainties remains to be performed. 

\section{The top quark mass}
\label{sec:mass}

The determination of the mass of the heaviest elementary is a key measurement, 
testing the relations that the Standard Model predicts between the $W$-boson
mass, the top quark mass and the Higgs boson mass (see, for 
instance, Ref.~\cite{Baak:2014ora}) and the stability of the vacuum 
at large scales~\cite{Degrassi:2012ry}.

The current world average for the top quark mass based on {\em direct} 
measurements at hadron colliders have attained a precision of better 
than 0.5\%~\cite{ATLAS:2014wva}. Several single measurements by CMS 
and D0 achieve 500~\mev{} uncertainty. After three years of operation
at approximately half the design energy the ATLAS and CMS already exceed the 
expectations~\cite{Beneke:2000hk} drawn up before the start of the LHC. 

Expectations for further progress at the LHC range from a pessimistic
500~\mev{}~\cite{Juste:2013dsa} after the complete LHC programme
to 200~\mev{}~\cite{CMS-PAS-FTR-13-017}. This uncertainty does not include 
the ambiguity in the interpretation, that is estimated to 
be ${\cal O} (1\gev)$~\cite{Juste:2013dsa,ahoang08,Moch:2014tta,ahoang14,Corcella:2015kth}. 
Significant theory progress
is required to make sure that the uncertainty in the interpretation of the measurements catches
up with experimental progress. A possible avenue is indicated by the method of 
Ref.~\cite{Butenschoen:2016lpz}, that may provide a means to {\em calibrate} 
the mass to a field-theoretical mass definition to a precision of several 100~\mev{}.
Given that the theory uncertainty, with an 
unclear future evolution, is likely to dominate the measurement at hadron colliders the 
FCChh Standard Model study group prefers not to give prospects for top quark 
mass measurements~\cite{Mangano:2016jyj}.

Measurements using alternative methods
with complementary sensitivity to the main systematics may reveal 
a possible tension not covered by the systematic uncertainties. Rigorous
pole mass extractions from (differential) cross sections are to 
reach 1~\gev{} precision in the near future~\cite{Alioli:2013mxa}.
Progress beyond this precision will depend on the development on new methods
and is therefore unpredictable.

\begin{table}[b!]
\caption{Brief summary of the prospects for the precision of the top quark mass measurement at different facilities. HL-LHC prospects are based on Ref.~\cite{CMS-PAS-FTR-13-017}. The experimental precision in an $e^+e^-$ threshold scan is based on Refs.~\cite{Seidel:2013sqa,Horiguchi:2013wra,Martinez:2002st} and the theory uncertainties on Refs.~\cite{Simon:2016htt,Vos:2016til}.}
\label{tab:mass}
\centering
\begin{tabular}{lx{2.5cm}x{2.5cm}x{6cm}} \\ \hline
\multicolumn{4}{c}{\bf top quark mass prospects} \\
project                      & $\sqrt{s} $  & $\int \cal{L}$      &      precision  \\ 
                             &     [\tev]        &   [\iab]                  &        [\mev]                      \\ \hline
Tev + LHC8                   &  1.96 + 8   & 0.01 + 0.02           &    500 (exp.) $\oplus$ 1000 (theo.) \\
HL-LHC                   & 14  & 3 &     200 (exp.) $\oplus$ ? (theo.) \\
$e^+e^-$ collider           & 0.35 & 0.1 & 20 (exp.) $\oplus$ 50 (theo.)   \\
new $pp$ collider    & 100~ & 20  &  ? (theo.) \\         \hline                
\end{tabular}
\end{table}

The cross section around the top quark pair production threshold 
at a lepton collider offers excellent sensitivity to the top quark mass~\cite{Gusken:1985nf}.
The shape can be predicted to good precision using a NNNLO NRQCD calculation~\cite{Beneke:2015kwa}.
The effect of Initial-State-Radiation, beam energy spread and beamstrahlung modify this shape somewhat, 
introducing slight differences between $e^+e^-$ collider projects. A fit to a ten-point threshold
scan with a total integrated luminosity of 100~\ifb{} yields a statistical precision on a threshold
mass (the 1S or PS mass) of approximately 20~\mev{}. The dominant uncertainty on the top quark mass
extraction is expected to be due to the theory uncertainty, in the prediction of the line 
shape~\cite{Simon:2016htt} and the conversion to a short-distance mass~\cite{Marquard:2015qpa} 
(such as the $\bar{MS}$ mass). Provided the strong coupling constant $\alpha_s$ is known
to sufficient precision the total theory uncertainty amounts to approximately 50~\mev{}.
The prospects for the top quark mass measurement of HL-LHC, lepton colliders and a future
high-energy $pp$ collider are summarized in Tab.~\ref{tab:mass}.


\section{Top quark QCD couplings}
\label{sec:qcd}

The QCD interactions of the top quark are already tightly constrained by 
measurements at the Tevatron and the LHC. A fit to all existing data provides
limits on the Wilson coefficients of dimension-six operators affecting the 
$t\bar{t}g$ vertex and the $q\bar{q} \rightarrow t\bar{t}$ four-fermion
operators that are order 0.1 $\times \Lambda^2/v^2$~\cite{Buckley:2015nca,Buckley:2015lku}. 
Measurements in extreme corners of phase space, where top
quarks are highly boosted and new techniques are required, provide
constraints of comparable strength to the more classical 
analyses~\cite{Englert:2016aei}. In the future such measurements are 
expected to become much more precise. With their much enhanced sensitivity
to the effect of new physics at high scales the constraints 
can be substantially improved~\cite{Rosello:2015sck}.

The potential of lepton colliders to constrain anomalous chromo-magnetic
moments was studied in Ref.~\cite{Rizzo:1994tu}, with results that are
competitive compared to the HL-LHC prospects.

The SPPC or FCChh are to take measurements on $t\bar{t}$ production
well beyond the energy reach of the LHC. Ref.~\cite{Aguilar-Saavedra:2014iga}
envisages the extraction of top-quark chromo-electric and chromo-magnetic
moments from a measurement of the cross section with $ m_{t\bar{t}} > $ 10~\tev.
Even with pessimistic assumptions on the ability of the experiments
to isolate highly boosted final states and a 5\% systematic uncertainty
such a measurement can improve the HL-LHC constraints by an order of magnitude.
The prospects for constraints on the top quark QCD interactions are summarized
in Tab.~\ref{tab:qcd}.

\begin{table}[h!]
\caption{Brief summary of the prospects for the precision of constraints on the top quark QCD interactions at different facilities. The expected 95\% C.L. limits on $d_V$ and $d_A$ for the Tevatron + LHC8, LHC14 and FCChh are taken from Ref.~\cite{Aguilar-Saavedra:2014iga}. The expected precision at lepton colliders is based on the old study of Ref.~\cite{Rizzo:1994tu}.    }
\label{tab:qcd}
\begin{tabular}{lx{2.5cm}x{2.5cm}x{6cm}} \\ \hline
\multicolumn{4}{c}{\bf top quark QCD coupling prospects $(d_V, d_A)$} \\
project                      & $\sqrt{s} $ [\tev] & $\int \cal{L}$ [\ifb]     &      expected precision    \\ \hline
Tev + LHC8                   &  1.96 + 8   & 0.01 + 0.02           &   $|d_V| < $ 0.02, $|d_A| <$ 0.09    \\
LHC                          & 14            & 0.3 &    $|d_V| < $ 0.01, $|d_A| <$ 0.02   \\
$e^+e^-$ collider           & \multicolumn{3}{c}{possibly competitive with HL-LHC at 1~\tev}             \\
new $pp$ collider                & 100     &  20  &  $|d_V|,|d_A| < $ 0.003 \\  \hline                       
\end{tabular}
\end{table}

\section{The top quark and the Higgs boson}
\label{sec:higgs}

A direct measurement of the interaction of the top quark and the Higgs boson
- with a unit Yukawa coupling - is a priority in high energy physics. 
ATLAS and CMS have some evidence that
associated $t\bar{t}H$ production indeed occurs and have $\cal{O}$ (100\%) 
uncertainty on the production rate. The precision of the measurement Yukawa
coupling is expected to reach $\cal{O}$ (10\%) at the end of the complete
HL-LHC programme~\cite{Dawson:2013bba}. 

The $t\bar{t}H$ final state is produced at lepton colliders. The cross section
shows a sharp turn on at around 500~\gev{} and reaches a broad maximum of
2 fb at about 800~\gev{}. Full-simulation studies have been performed
at several center-of-mass energies between 0.55~\tev{} and 1.5~\tev{}. 
The top quark Yukawa coupling can be determined to 3-4\% precision in this 
interval~\cite{Abramowicz:2016zbo,Fujii:2015jha,Price:2014oca} with the
integrated luminosities envisaged for the ILC and CLIC.

At a 100~\tev{} hadron collider the $t\bar{t}H$ production cross section
is over 30~pb. The (NLO) theory uncertainty on the predicted rate is 
approximately 10\%, dominated by scale uncertainties. Ref.~\cite{Plehn:2015cta}
shows that in the ratio of rates for two closely related associated production 
processes, $t\bar{t}H$ and $t\bar{t}Z$, the uncertainty is reduced to
1-2\%. The robustness of uncertainty estimate based on the (synchronous)
variation of the renormalization and factorization scales in ratios of
cross sections can be tested at the LHC (cf. the ATLAS measurement
of the ratio of the $t\bar{t}$ cross sections at 7~\tev{} and 
8~\tev{}~\cite{Aad:2014kva} or the ratio of top and Z cross 
sections~\cite{ATLAS-CONF-2015-049}). 
A phenomenological study of mildly boosted $t\bar{t}H$ events 
in the same paper predicts that an extraction of the Yukawa coupling
to 1\% precision is feasible. While a detailed experimental study is
still lacking, it is clear that a high-energy hadron collider is
the only machine on the horizon with the potential to reach this precision.
The prospects of all future facilities are summarized in Tab.~\ref{tab:higgs}.

\begin{table}[h]
\caption{Brief summary of the prospects for the precision of measurements of the top quark Yukawa coupling at different facilities. HL-LHC prospects are taken from Ref.~\cite{Dawson:2013bba}. The expected precision at a linear lepton colliders incorporates the results of several ILC and CLIC studies at center-of-mass energies ranging from 0.55-1.5~\tev{} and assumed integrated luminosities ranging from 1-4~\iab{}~\cite{Fujii:2015jha,Abramowicz:2016zbo,Price:2014oca}. FCChh prospects are taken from Ref.~\cite{Plehn:2015cta}.    }
\label{tab:higgs}
\begin{tabular}{lx{2.5cm}x{2.5cm}x{6cm}} \\ \hline
\multicolumn{4}{c}{\bf top quark Yukawa coupling prospects} \\
project                      & $\sqrt{s}$ [\tev] & $\int \cal{L}$ [\iab]     &      expected precision    \\ \hline
HL-LHC                   & 14~\tev{}  & 3 &    10\%   \\
$e^+e^-$ collider           & 0.55-1.5~\tev{} & 1-4 & 3-4\% \\
$pp$ collider    & 100~\tev{} & 10-20  &  1\% \\   \hline 
\end{tabular}
\end{table}

\section{Top quark FCNC interactions}
\label{sec:fcnc}

Flavour changing neutral current (FCNC) interactions involving top quarks 
are suppressed to well below detectable levels in the Standard Model, but
can reveal sizeable contributions from a variety of new physics setups.
The limits on the strenght of the $tq\gamma$, $tqZ$, $tqH$ and 
$tqg$ vertices from LEP, HERA and the Tevatron are rapidly being
superseded by LHC searches for rare top decays~\cite{Durieux:2014xla}. 

Future lepton colliders may probe FCNC interactions in top quark production
through $e^+e^- \rightarrow tj$ production~\cite{AguilarSaavedra:2000db} or 
in decays~\cite{Zarnecki:2016bgr}. With the relatively small top quark pair 
production rate compared to hadron colliders, the latter is competitive 
primarily for channels that are particularly challenging at the LHC. 
An example is  $ t \rightarrow c H$ that can be constrained to 
$BR \sim 10^{-5}$ by the ILC at $\sqrt{s} = $ 500~\gev{}~\cite{Zarnecki:2016bgr}, 
compared to $BR \sim 10^{-4}$ for the HL-LHC programme~\cite{Agashe:2013hma}.

Hadron colliders at $\sqrt{s} = $ 100~\tev{} produce billions of top quark and
 can potentially limit the branching ratios for rare top quark decays such as 
$t \rightarrow q\gamma$ and $tqZ$ to the $10^{-7}$ level (two orders
better than HL-LHC). Dedicated studies are missing, however, and 
Ref.~\cite{Mangano:2016jyj} warns that ``the systematic uncertainties in the
background predictions will likely be dominant, and a more reliable 
estimation of the sensitivity requires a detailed analysis.” A brief summary
of the FCNC prospects of the different projects is presented in Tab.~\ref{tab:fcnc}.

\begin{table}[h]
\caption{Brief summary of the prospects for the precision of measurements of the top quark Yukawa coupling at different facilities. HL-LHC prospects are taken from Ref.~\cite{Dawson:2013bba}, with more recent additions from Ref.~\cite{ATL-PHYS-PUB-2016-019}. The expected precision at a linear lepton collider is studied in Ref.~\cite{Vos:2016tof,Zarnecki:2016bgr}. A discussion of the potential of the FCC is found in Ref.~\cite{Mangano:2016jyj}  }
\label{tab:fcnc}
\begin{tabular}{lx{2.5cm}x{2.5cm}x{6cm}} \\ \hline
\multicolumn{4}{c}{\bf top quark FCNC interaction prospects} \\
project                      & $\sqrt{s}$ [\tev] & $\int \cal{L}$ [\iab]     &      expected precision    \\ \hline
HL-LHC                   & 14~\tev{}  & 3 &    $BR(t\rightarrow qX) \sim 10^{-5} - 10^{-4}$    \\
$e^+e^-$ collider           & 0.5~\tev{} & 4 & $BR(t\rightarrow cH) \sim 10^{-5}$ \\
$pp$ collider    & 100~\tev{} & 10-20  &  possibly $BR(t\rightarrow qX) \sim 10^{-7}$ \\   \hline 
\end{tabular}
\end{table}

\section{Top quark EW couplings}
\label{sec:ew}

The EW couplings of the top quark, in particular those involving neutral
gauge bosons, are among the least constrained parameters of the Standard
Model. Models with a new strong sector (Randall Sundrum, composite Higgs)
tend to predict large deviations from the SM 
predictions~\cite{Richard:2014upa}. 

The HL-LHC can constrain the $t\bar{t}\gamma$ and $t\bar{t}Z$ vertices directly
by studying associated production. In run I these processes were observed
and the production cross sections measured to $\cal{O}$ (30\%). Measurements
of the $Wtb$-vertex from studies of top quark decay and single top production 
can be used to derive constraints on the same operators in an EFT 
analysis~\cite{Buckley:2015nca}.

The relatively rare associated production processes benefit strongly from an
increase in the center-of-mass energy, with statistical uncertainties
becoming irrelevant. Recent studies~\cite{Rontsch:2014cca,Schulze:2016qas} 
show that the (NLO) theory uncertainty can largely be circumvented by forming
the ratio of $t\bar{t}X / t\bar{t}$, where $X$ is a $Z$-boson or a photon.
A simulteanous measurement of both ratios strongly improves
the limits on EW magnetic and electric dipole moments.

Future lepton colliders with sufficient energy to produce top quark pairs
are the natural place to study $t\bar{t}\gamma$ and $t\bar{t}Z$ 
vertices~\cite{Amjad:2013tlv,Amjad:2015mma,Janot:2015yza,Khiem:2015ofa}. 
Constraints on the form factors reach sub-\% precision, one to two orders
beyond the prospects of the HL-LHC and well beyond the reach of even
a 100~\tev{} hadron collider.

\begin{table}[h]
\caption{Brief summary of the prospects for the precision of constraints on the top quark QCD interactions at different facilities. Results are presented as expected 68\% C.L. limits on the form factors of the $t\bar{t}Z$ vertex. The expected precision at HL-LHC is based on Ref.~\cite{Baur:2004uw}, that at lepton colliders on Ref.~\cite{Amjad:2015mma}.    }
\label{tab:ew}
\begin{tabular}{lx{2.5cm}x{2.5cm}x{6cm}} \\ \hline
\multicolumn{4}{c}{\bf top quark EW coupling prospects $(F_{1V},F_{2V},F_{1A},F_{2A})$} \\
project                      & $\sqrt{s} $ [\tev] & $\int \cal{L}$  [\iab]    &      expected precision    \\ \hline
HL-LHC                   & 14~\tev{}  & 300~\ifb{} &   $\sim$ 0.03-0.2   \\
$pp$ collider    & 100~\tev{} & 10-20~\iab{}  &  factor 3-10 wrt HL-LHC \\     
$e^+e^-$ collider           & 0.5~\tev{} &  500~\ifb{} & $\sim$ 0.002-0.005   \\   \hline
\end{tabular}
\end{table}

\section{Summary and outlook}
\label{sec:summary}

New energy-frontier facilities offer new opportunities for top physics
in the post-LHC era. In this contribution I have reviewed the potential
of future lepton and hadron colliders, with a focus on measurements with 
exquisite BSM sensitivity. 

An $e^+e^-$ collider with a center-of-mass energy exceeding $ 2 m_t $ enables 
the scrutiny of the top quark in a precision environment. An energy scan
through the top quark pair production threshold can realistically 
yield a determination of the $\bar{MS}$ mass to 50~\mev{}.
The comparison of per-mil-level measurements with equally precise SM 
predictions provides constraints on the $t\bar{t}Z$ and $t\bar{t}\gamma$
vertices that are an order of magnitude more stringent than those expected
at the HL-LHC, indirectly probing new physics scales in excess of 10~\tev. 
A direct measurement of 
the top quark Yukawa coupling in associated $t\bar{t}H$ production can
be performed to a precision of 3-4\% at (linear) colliders with a 
center-of-mass energy exceeding 550~\gev.

A very large hadron collider, with a center-of-mass energy up to 100~\tev,
will extend the mass reach of direct searches and provide very tight
constraints on the QCD interactions of the top quark. The FCChh has 
the potential to measure the top quark Yukawa coupling to 1\% precision. 
In many areas, such as the top quark mass measurement and 
searches for FCNC decays such a machine has clear potential to go
well beyond the state-of-the-art in 2037, but thorough studies are
yet to be performed. 

Much work remains to be done to establish a complete picture of the
top quark physics potential of different facilities. Many new studies 
are expected to appear before the European strategy update, that is
to conclude in the first half of 2020.
Progress in theory and new results at the LHC may tell us what level of 
control to expect over systematic uncertainties from the construction 
of carefully chosen cross-section ratios.
The emerging global EFT analyses can provide a valuable tool to interpret 
the BSM sensitivity of measurements in different processes, at different
energy and with different precision.

\Acknowledgements
The results in this contribution are the result of hard work of a small group 
of people devoted to the long-term future of this field.


\bibliographystyle{JHEP}
\bibliography{top}


\end{document}